# Unravelling the role of the interface for spin injection into organic semiconductors

Clément Barraud[1], Pierre Seneor[1*], Richard Mattana[1*], Stéphane Fusil[1], Karim Bouzehouane[1], Cyrile Deranlot[1], Patrizio Graziosi[2], Luis Hueso[2], Ilaria Bergenti[2], Valentin Dediu[2], Frédéric Petroff[1], Albert Fert[1]

[1]*Unité mixte de Physique CNRS/Thales, 91767 Palaiseau France, and Université Paris-Sud, 91405 Orsay, France*

[2]*ISMN-CNR, Via Gobetti 101, 40129 Bologna, Italy*

**Whereas spintronics brings the spin degree of freedom to electronic devices, molecular/organic electronics adds the opportunity to play with the chemical versatility. Here we show how, as a contender to commonly used inorganic materials, organic/molecular based spintronics devices can exhibit very large magnetoresistance and lead to tailored spin polarizations. We report on giant tunnel magnetoresistance of up to 300% in a $(La,Sr)MnO_3/Alq_3/Co$ nanometer size magnetic tunnel junction. Moreover, we propose a spin dependent transport model giving a new understanding of spin injection into organic materials/molecules. Our findings bring a new insight on how one could tune spin injection by molecular engineering and paves the way to chemical tailoring of the properties of spintronics devices.**

Molecular spintronics, by combining the potential of spintronics[1] and molecular/organic electronics [2], is now considered as a promising alternative to conventional spintronics with inorganic materials[3]. Besides chemical flexibility and low production costs, the opportunity that spin relaxation times could be enhanced by several orders of magnitude compared to inorganic materials arouse a strong interest for organic semiconductors (OSC)[4]. Weak spin-orbit coupling associated to light element compounds and electronic transport via $\pi$ delocalized orbitals would be involved in explaining this spectacular gain of spin lifetime. The pioneer spin-valve effects were observed by V. Dediu *et al.* [5]. for sexithiophene. Since then, most of the studies have been focused on $Alq_3$ (tris[8-hydroxyquinoline]aluminum)[6-13] motivated by the achievement of sizeable inverse spin valve effect by Z.H. Xiong *et al.*[6]. However, the mechanisms underlying the spin injection into the OSCs are still to be unravelled and remain one of the main challenge of this new uprising field[14]. Here, we report on giant tunnel magnetoresistance (MR) up to 300% in $(La,Sr)MnO_3/Alq_3 \sim 2nm/Co$ nanojunctions. We further develop a spin dependent transport model giving an understanding of spin injection into organic materials/molecules and opening new opportunities for chemically tailored spintronics devices. Ultimately, thanks to molecular engineering, the physical properties of spintronics devices could be expected to be tailored through playing with the anchoring groups and the backbone of the molecules.

The $LSMO/Alq_3/Co$ tunnel junctions are elaborated from $LSMO/Alq_3$ (10-30nm) bilayers using a conductive tip AFM nanolithography (CT-AFM) process. The sample is nanofabricated in order to circumvent problems such as inhomogeneity and strong metal diffusion[6] that would lead to inevitable short-circuits in wide area tunnel junctions. The desired Alq$_3$ tunnel barrier thickness left after the indentation is obtained by controlling the AFM tip penetration into the layer. The indentation process is triggered by the tunnel current between the tip and the sample. Finally, the nanoholes are filled with Co leading to a LSMO/Alq$_3$/Co magnetic tunnel junction device (see Fig. 1. and methods).

As a preamble, in order to characterize the Alq$_3$ thickness versus tip to sample resistance, we defined matrices of nanoindents in which sequences of resistances in the range of $10^{5.5}\Omega$- $10^{11}\Omega$ were taken as threshold values to stop the process. The lower bound corresponds to the LSMO/tip contact resistance measured on a free standing surface of LSMO. A reference nanoindent realized at this threshold resistance value is presented in Fig. 2a. As an example, on the same figure, we show another nanoindent realized at $10^{9.5}$ $\Omega$ and leaving 4 nm of Alq$_3$ tunnel barrier. Furthermore, the thickness of the remaining organic barrier is presented as a function of the preset resistance threshold in Fig. 2b. The linear increase of the resistance versus barrier thickness in logarithmic scale unravels the exponential character of the tunneling mechanism. Fitting the exponential increase with $G \propto e^{-\beta d}$ (where d is the barrier thickness[2]) we find a value of $\beta = 0.23$ Å$^{-1}$ which is in agreement with $\beta \approx 0.3$ Å$^{-1}$ extracted from tunneling measurements in MTJs[13]. Nevertheless, this corresponds to a much slower decay for the electronic wave function in Alq$_3$ MTJs compared to oligothiophene[15] ( $\beta = 0.41$ Å$^{-1}$), alkanethiol SAMs[2] ($\beta \approx 0.8 - 1.08$ Å$^{-1}$) or inorganic MgO based[16] ($\beta \approx 0.7$ Å$^{-1}$) tunnel junctions. A simple picture of this small $\beta$ can be given in terms of tunnel barrier height. It was shown that the dipole formation at the metal/organic interface shifts the organic level towards lower energies of 0.9 eV for LSMO/Alq$_3$ [17], resulting in an expected mean electron barrier height of 1.3 eV. Furthermore, for thin organic films in the nm range, a significant reduction of the effective barrier height is expected due to the image potential contribution[18]. Indeed, the reduced dielectric constant ($\epsilon$) in $Alq_3$ barriers ($\epsilon_{Alq3}$ = 1.6 [19]) compared to conventional $Al_2O_3$ inorganic barriers ($\epsilon_{Al2O3}$ = 10) enhances the potential image effect by the ratio $\epsilon_{Al2O3}/\epsilon_{Alq3}$. This reduces the effective mean barrier height by at least several hundreds meV for thin layers. For example, according to ref. 18 a reduction of 1 eV can be expected for a 2 eV and 2 nm thick tunnel barrier.

In the following, we present the spin dependent transport measurements performed on the LSMO/$Alq_3$ ~ 2nm/Co nanojunctions. The current-voltage curves (inset of Fig. 3) are clearly non-linear confirming a tunnel transport behaviour through the $Alq_3$ organic barrier as expected from the resistance versus $Alq_3$ thickness described in Fig. 2. The I-V curves show some fine structures at 2 K that fade with increasing temperature. These structures could originate from phonons or magnons (especially at low bias voltage below 100 mV[20]), extrinsic (structural defects, atoms...) and/or intrinsic electronic states in the gap. Here, we can reasonably rule out Coulomb blockade through a single small metallic cluster as the origin of this low bias fine structure (below 100 mV). Indeed, taking into account the low dielectric constant of organic compounds such as $Alq_3$, fine structure on a voltage scale lower than 100 mV would require metallic clusters with diameters larger than 7 nm [21]. This is much larger than the organic barrier thickness of the sample.

The variation of the resistance as a function of the in-plane magnetic field recorded at 2K and -5 mV is shown in Fig. 3. The observed magnetoresistance reaches 300%. Clear and reproducible positive magnetoresistance effects have been obtained at low temperature and at different bias voltage (Fig. 4a). The lower coercive field is ascribed to the LSMO electrode (confirmed by SQUID measurements) and the higher to the Co nanocontact. SQUID measurements performed on LSMO/Alq$_3$ bilayers reveal also a low remanent magnetization (Mr/Ms ≈ 0.4). This low remanence induces a higher resistance at zero magnetic fields than at high magnetic field where the magnetization of LSMO is fully saturated. Concerning the MR origin, we have checked by angular dependence measurements that the measured effects were not due to tunneling anisotropic magnetoresistance. In the inset of Fig. 4b, we show the MR temperature dependence. The MR undergoes a sharp decrease in temperature with only 25% remaining at 120 K and below noise level at 180 K. The decrease of the MR with temperature cannot be ascribed only to the rapid loss of magnetization at the LSMO surface [22]. It also indicates a loss of spin polarization during the tunnel transport. This is supported by the bias voltage dependence of the magnetoresistance (Fig. 4b) obtained from magnetoresistance measurements recorded at different bias voltage and from I-V curves both in parallel (PA) and antiparallel (AP) magnetic configurations. We observe a strong decrease of magnetoresistance at low voltage followed by a slower decrease at higher voltage. Note that the MR remains positive whatever the applied bias voltage. Again, the decrease of magnetoresistance is stronger than the one observed in LSMO based inorganic tunnel junctions[22] explained by magnon excitations[20]. In our organic tunnel junctions the MR is reduced by a factor 2 at 25 mV (see Fig. 4b) and the magnetoresistance effect vanishes at 200 mV. We observe a correlated magnetoresistance and resistance decrease (not shown here) in temperature and voltage. In addition to magnons which are known to play an important role in MTJs, phonons are also expected to exert a key influence for organic barriers[23]. For instance, it has been demonstrated that phonons play a significant role in the coupling of certain molecules to a metallic surface[24].

We first discuss on the amplitude of the magnetoresistance. Assuming that the spin polarization of the LSMO/Alq$_3$ interface (P$_{LSMO}$) is positive and fully polarized[25], the spin polarization of the Co/Alq$_3$ interface, obtained from the simple Jullière's formula TMR=$(R_{AP}-R_{PA})/R_{PA}$ =$2P_{Co}P_{LSMO}/(1-P_{Co}P_{LSMO})$, reaches at least + 60%. Note that this ideal situation for LSMO corresponds to a lower bound for the cobalt spin polarization. A lower LSMO spin polarization would lead to a higher Co spin polarization. The positive sign of P$_{Co/Alq3}$ in our MTJs is in agreement with the results of T.S. Santos *et al.*[13]. They extract a positive sign for P$_{Co/Alq3}$ (+ 27%) from Meservey-Tedrow measurements[26] in Co/Alq$_3$/Al tunnel junctions. Although we obtain the same sign, the magnitude is much higher in our junctions. This could be explained by pointing out that in contrast to commonly measured large area MTJs (micron size and above) we probe only local nanometer scale properties. This is of particular relevance for highly inhomogeneous layers such as organic thin films. The high spin polarization can also be linked to specific spin dependent hybridization of the orbitals at the metal/organic interface[29]. For example, chemical reaction and complex formation at the interface between the electrode and Alq$_3$ were proposed by A.N. Caruso *et al.*[30] as a possible mechanism to explain the shift of the band structure of Alq$_3$ deposited on various non magnetic metallic electrodes.

We now discuss on the sign of the magnetoresistance of LSMO/Alq$_3$/Co trilayers. Since the pioneer result of Z.H. Xiong *et al.*[6], inverse spin valve effect has been regularly observed in the case of mm large and thick[7-10] Alq$_3$ layers ($\approx$ 130 - 250 nm[31]). It is commonly accepted to analyze such sign in the framework of the Jullière model leading to opposites spin polarizations for the interfaces. With LSMO (P$_{LSMO}$ > 0) thought of as a spin analyzer[25], this naturally lead to P$_{Co/Alq3}$ < 0. However, it was pointed out that both positive and negative MR could be observed in thick LSMO/Alq$_3$/Co devices as a function of applied voltage[12]. In addition, for the same LSMO/Alq$_3$ bilayers, here we report a positive MR for locally probed thin tunnel barriers while a negative one has been observed for larger and thicker ones[11]. While one could relate those last discrepancies to materials fluctuations between studies, in the following we propose a description of a spin injection mechanism explaining the observed discrepancies.

We now show how the formation of localized states in the first molecular layer at the electrode interface can change completely the MR of organic spin valves with respect to what is usually found in conventional inorganic ones. This can lead to an increase of the effective spin polarization of the electrodes or even change their sign. For this purpose we introduce a simple 1D model following Bardeen's approach. For a system with transmission $T_{if}^{\sigma\sigma'}$ depending on the initial $i$ and final state $f$ and spin directions $\sigma$, one can write the conductance at zero bias as:

$$\text{G}^{\sigma\sigma'}(\text{E}_\text{F}) = \frac{2\pi\text{e}^2}{\text{h}} \sum_{(i,f)} \left|\text{T}_{if}^{\sigma\sigma'}\right|^2 \delta(\text{E}_i\text{-E}_\text{F})\delta(\text{E}_f\text{-E}_\text{F}) \quad (1).$$

Usually, in the case of direct tunneling, the transmission coefficient is kept constant and the conductance is rewritten as proportional to the spin dependent density of states (DOS) of the electrodes $N^{\sigma}(E)$. This leads to the original Jullière formula for the TMR with the spin polarization expressed as $P = (N^{\uparrow} - N^{\downarrow})/(N^{\uparrow} + N^{\downarrow})$. In a more realistic approach, the transmission coefficient is also affected by the specific bonding of different band states of the electrodes at the organic interface, which leads to a weighting of the DOS contribution. This gives rise to an "effective" DOS entering the spin polarization.

Recalling that most of the OSCs such as Alq$_3$ are closer to small molecules than conventional semiconductors, the first molecular layer is thought to have a key role for charge injection into such OSCs[32]. In the following we describe how the first molecular state may contribute to this effective spin dependent DOS. For a ferromagnetic metal, one could expect that a spin dependent broadening (spectral density $\Gamma$) of those localized states arises from the coupling to the electrode. This gives for the left electrode (L) and spin orientation ($\sigma = \uparrow, \downarrow$):

$$\Gamma_L^\sigma(E) = 2\pi \sum_i \left| V_{Di}^\sigma \right|^2 \delta(E_i - E_F) \quad (2),$$

where $V_{Di}^\sigma$ is the coupling energy between the state i of the left electrode and the donor state D represented in Fig. 5a. A similar equation holds for the acceptor state at the right electrode. This spectral density can be seen as a weighted DOS for the transmission from the left electrode to the D state. For constant coupling, we obtain $V_{Di}^\sigma \approx V$ and $\Gamma^\sigma(E) \propto N^\sigma(E)$. Following the adapted concept of bridges developed in the scattering theory for molecular junctions[33,34], the transmission of equation (1) can be expressed as:

$$T_{if}^{\sigma\sigma'} = \frac{V_{iD}^\sigma}{\Delta \tilde{E}_D^\sigma + i \, \Gamma_L^\sigma/2} T_{DA} \frac{V_{Af}^{\sigma'}}{\Delta \tilde{E}_A^{\sigma'} + i \, \Gamma_R^{\sigma'}/2} \quad (3),$$

where $\Delta \tilde{E}_D^\sigma$ (resp. $\Delta \tilde{E}_A^{\sigma'}$) is the donor (resp. acceptor) energy difference with respect to the Fermi level. The tilde highlights that this energy difference includes the two main contributions introduced above: disorder in the interfacial dipole fields and images forces[32], and a shift due to the real part of the self-energy induced by the coupling. $T_{DA}$ is the bulk transmission in the OSC between the donor and acceptor states as shown in Fig. 5a. This transmission factor is not the object of this study but could represent different transport conditions from superexchange to hopping.

Two limits can be described depending on whether $\left|\Gamma\right|>>\left|\Delta\tilde{E}\right|$ or $\left|\Gamma\right|<<\left|\Delta\tilde{E}\right|$. The first case is likely to happen for strong metal/molecule coupling (large $\Gamma$). For example, broadening in the eV range[35] have been predicted. The second case corresponds mainly to a weak interaction for which the spin dependent $\Gamma$ is low enough to be neglected. It is to be expected that this interface related interaction will depend on the selected metal/molecule couple and its deposition condition.Accordingly, a metal deposited on a molecule surface or a molecule on a metal surface could lead to different couplings. As far as $\Delta E$ is concerned, two effect will lead to its reduction: the metal/molecule coupling and the distribution of image forces and interfacial dipole fields. The latter one has been shown by Baldo and Forrest[32] to have a leading role for charge injection into the organic as the tail of the distribution brings states close to the metal's Fermi level.

Below we illustrate these two limits. The conductance can be expressed from equations (2) and (3) as:

$$G^{\sigma\sigma'}(E_F)=\frac{2e^2}{h}\left|T_{DA}\right|^2\gamma_L^\sigma(E_F)\gamma_R^{\sigma'}(E_F) \quad (4),$$

where $\gamma^\sigma(E_F)=1/\Gamma^\sigma(E_F)$ in the $\left|\Gamma\right|>>\left|\Delta\tilde{E}\right|$ limit (Fig. 5b) and $\gamma^\sigma(E_F)=\frac{\Gamma^\sigma(E_F)}{\left|\Delta\tilde{E}^\sigma\right|^2}$ in the $\left|\Gamma\right|<<\left|\Delta\tilde{E}\right|$ limit. The striking point is that while the conductance is still proportional to the spectral density (as usually expressed for direct tunneling) in the weakest coupling regime it is now inversely proportional to the spectral density in the strongest coupling regime. This leads to a Jullière like formula with an effective spin polarization ($P^*$) *inverted* for $\left|\Gamma\right|>>\left|\Delta\tilde{E}\right|$ and *levered* for $\left|\Gamma\right|<<\left|\Delta\tilde{E}\right|$:

$$TMR=\frac{2P_L^*P_R^*}{1-P_L^*P_R^*} \qquad (5),$$

with $P^* = -\dfrac{\Gamma^\uparrow - \Gamma^\downarrow}{\Gamma^\uparrow + \Gamma^\downarrow}$ in the $\left|\Gamma\right| >> \left|\Delta\tilde{E}\right|$ limit (6)

and $P^* = \dfrac{\Gamma^\uparrow / \Delta\tilde{E}^\uparrow - \Gamma^\downarrow / \Delta\tilde{E}^\downarrow}{\Gamma^\uparrow / \Delta\tilde{E}^\uparrow + \Gamma^\downarrow / \Delta\tilde{E}^\downarrow}$ in the $\left|\Gamma\right| << \left|\Delta\tilde{E}\right|$ limit (7).

As a consequence, one could use the metal/molecule coupling as a new way to tailor the properties of spintronics devices by adequately combining ferromagnetic metals to molecules and/or their anchoring groups. For intermediate couplings however, the local effect of image forces and dipole field disorder reducing the value of $\Delta\tilde{E}$ will play a significant role.

This model explains the spin injection into molecules and all the apparent discrepancies6-13 on the observed MR signs in LSMO/Alq$_3$/Co spin valves discussed above. Following the Jullière model (and using LSMO as a spin analyzer[25] with $P^*_{LSMO/Alq3} > 0$) $P^*_{Co/Alq3} < 0$ at the Co/Alq$_3$ interface is necessary to explain the commonly observed negative MR sign in large area junctions[6-10]. However, we propose that, oppositely, an inversion occurs in large area samples at the LSMO/Alq3 interface leading to an effective $P^*_{LSMO/Alq3} < 0$ while having $P^*_{Co/Alq3} > 0$. The $P^*_{Co/Alq3} > 0$ sign, suggesting a weak coupling in the model, is supported by two experimental results: i) $P_{Co/Alq3} > 0$ was directly measured by Meservey-Tedrow technique[13]; ii) inserting an inorganic Al$_2$O$_3$ spacer between Alq$_3$ and Co in a LSMO/Alq$_3$/Co spin valves did not change the sign of the MR[11] whereas an inversion would have been expected due to commonly reported $P_{Co/Al2O3} > 0$. The $P^*_{LSMO/Alq3} < 0$ sign would suggest a large enough intermediate coupling in the model in addition to disorder contribution. The disorder contribution being statistically distributed (a width of around 0.4 eV width has been shown for a range of cathodes on Alq3) the inversion should not happen for all the interfacial molecular states but only locally for those brought close enough to the electrode's Fermi level dominating the injection step[32]. The local character of this inversion explains the observation of $P^*_{LSMO/Alq3}$ >0 in our solid state SP-STM like experiment where we only probe a single outcome of the energetic disorder distribution and mainly remain off resonance. This is well emphasized by the results of Vinzelberg et al.[12] who mainly observe negative MR signs on mm wide LSMO/Alq3/Co spin valves while still reporting positive MR that inverse with bias voltage for few samples. A similar MR inversion as a function of bias has already been reported for spin dependent tunnelling through single localized states (corresponding to a merged donor-acceptor level in the model) in inorganic MTJs[36-37].

This shows that beyond the now recognized low cost, flexibility and long spin lifetime interest, chemistry could bring new properties hardly available in conventional inorganic spintronics. The unravelled specific spin dependent injection mechanism could be used to tailor the properties of spintronics devices through metal/molecule coupling. One particularly interesting new way would be to use self-assembled monolayers that could be grafted on any surface and where one could easily play with the anchoring groups and the backbone of the molecules, ultimately controlling the coupling by the applied voltage.

## <u>Methods:</u>

Fabrication nanometric size magnetic tunnel junctions $LSMO/Alq_3/Co$ is performed from $LSMO/Alq_3$ bilayer. The $Alq_3$ organic spacer in the 10nm-30nm range is grown by organic molecular beam epitaxy on a $La_{0.7}Sr_{0.3}MnO_3$ bottom electrode[11]. Films show a typical peak to peak roughness of 5 nm with an average hole to peak distance of hundred nanometers. As expected, it prevents the reliable fabrication of nanometer thick tunnel junctions free of short-circuits over micron size areas. We subsequently elaborate these nano-sized tunnel junctions with controlled organic barrier thickness in the nanometer range using a conductive tip AFM (CT-AFM) based nanofabrication process[38]. We start by capping the Alq$_3$ with a 30 nm thick resist protective layer and proceed to the nanoindentation. The real time monitoring of the conductivity between the tip and the LSMO electrode during the indentation process allows us to control the barrier thickness with precision. The conductive tip-LSMO resistance decreases by six orders of magnitude over 6nm while digging through the semiconducting organic layer down to the LSMO electrode. The organic barrier thickness left after the indentation is defined by stopping the nano-indentation process at a preset threshold resistance value. Here, the nanohole section is limited by the local radius at the tip end which is less than 10nm. The last step is to fill by sputtering the nano-hole with a cobalt ferromagnetic layer acting as a top electrode.

[1] Fert, A. Nobel Lecture: Origin, development, and future of spintronics. *Rev. Mod. Phys.* **80**, 1517-1530 (2008)

[2] Cuniberti, G., Richter, K. & Fagas, G. (eds.) Introducing Molecular Electronics, vol. 680 (Spinger Verlag, Berlin, 2005).

[3] Dediu, V., Hueso, L.Bergenti, I & Taliani, C. Spin routes in organic semiconductors *Nature Mat.,* **8**, 707 (2009)

[4] Sanvito, S. & Rocha, A. R. molecular spintronics: The art of driving spin through molecules. *J. Comput. Theor. Nanosci.* **3**, 624–642 (2006).

[5] Dediu, V., Murgia, M., Matacotta, F. C., Taliani, C. & Barbanera, S. Room temperature spin polarized injection in organic semiconductor. *Solid State Commun.* **122**, 181 (2002).

[6] Xiong, Z. H., Wu, D. & Vardeny, Z. V. Giant magnetoresistance in organic spin-valves. *Nature* **427**, 821 (2004).

[7] Majumdar,S., Majumdar, H.S., Laiho, R. & Osterbacka, R., Comparing small molecules and polymer for future organic spin-valves. *J. Alloys Compd.* **423** 169–171 (2006).

[8] Wang, F. J., Yang, C. G., Vardeny, Z. V. & Li, X. Spin response in organic spin valves based on $La_{2/3}Sr_{1/3}MnO_3$ electrodes. *Phys. Rev. B* **75**, 245324 (2007).

[9] Xu, W. et al. Tunneling magnetoresistance observed in $La_{2/3}Sr_{1/3}MnO_3$/organic molecule/Co junctions. *Appl. Phys. Lett.* **90**, 072506 (2007).

[10] Hueso, L. E., Bergenti, I., Riminucci, A., Zhan, Y. Q. & Dediu, V. Multipurpose magnetic organic hybrid devices. *Adv. Mater.* **19**, 2639 (2007).

[11] Dediu, V. et al. Room-temperature spintronic effects in $Alq_3$-based hybrid devices. *Phys. Rev. B* **78**, 115203 (2008).

[12] Vinzelberg, H. et al. Low temperature tunneling magnetoresistance on $(La,Sr)MnO_3$/Co junctions with organic spacer layers. *J. Appl. Phys.* **103**, 093720 (2008).

[13] Santos, T. S. et al. Room-temperature tunnel magnetoresistance and spin-polarized tunneling through an organic semiconductor barrier. *Phys. Rev. Lett.* **98**, 016601 (2007).

[14] Editorial, Why going organic is good *Nature Mat.,* **8**, 691 (2009)

[15] Sakaguchi, H. et al. Determination of performance on tunnel conduction through molecular wire using a conductive atomic force microscope. *Appl. Phys. Lett.* **79**, 3708 (2001).

[16] Matsumoto, R. et al. Oscillation of giant tunneling magnetoresistance with respect to tunneling barrier thickness in fully epitaxial Fe/MgO/Fe magnetic tunnel junctions. *Appl. Phys. Lett.* **90**, 252506 (2007).

[17] Zhan, Y. Q. et al. Alignment of energy levels at the $Alq_3$/ $La_{0.7}Sr_{0.3}MnO_3$ interface for organic spintronic devices . *Phys. Rev. B* **76**, 045406 (2007).

[18] Akkerman, H. B. et al. Electron tunneling through alkanedithiol self-assembled monolayers in large-area molecular junctions. *Proc. Natl Acad. Sci. USA* **104**, 11161 (2007).

[19] Vàzquez, H., Flores, F. & Kahn, A. Induced density of states model for weakly interacting organic semiconductor interfaces. *Org. Elec.* **8**, 241 (2007).

[20] Bratkovsky, A. M. Assisted tunneling in ferromagnetic junctions and half-metallic oxides. *Appl. Phys. Lett.* **72**, 2334 (1998).

[21] Bernand-Mantel, A. et al. Evidence for spin injection in a single metallic nanoparticle: A step towards nanospintronics . *Appl. Phys. Lett.* **89**, 062502 (2006).

[22] Bowen, M. et al. Observation of Fowler-Nordheim hole regime across an electron tunnel junction due to total symmetry filtering. *Phys. Rev. B* **73**, 140408(R) (2006).

[23] Wang, W. & Richter, C. A. Spin-polarized inelastic electron tunneling spectroscopy of a molecular magnetic tunnel junction. *Appl. Phys. Lett.* **89**, 153105 (2006).

[24] Tautz, F. S. et al. , Strong electron-phonon coupling at a metal/organic interface: PTCDA/Ag(111), *Phys. Rev. B* **65**, 125405 (2002)

[25] DeTeresa, J. M. et al. Role of the metal-oxide interface in determining the spin polarization of magnetic tunnel junction. *Science* **286**, 507 (1999).

[26] Meservey, R. & Tedrow, P. M. Spin-polarized electron-tunneling . *Physics Reports* **238**, 173 (1994).

[27] Alvarado, S. F., Rossi, L., Müller, P., Selder, P. F. & Riess, W. STP-excited electroluminescence and spectroscopy on organic materials for display applications. *IBM J. Res. & Dev.* **45**, 98 (2001).

[28] Yuasa, S., Fukushima, A., Kubota, H., Suzuki, Y. & Ando, K. Giant tunneling magnetoresistance up to 410% at room temperature in fully epitaxial Co/MgO/Co magnetic tunnel junctions with bcc Co(001) electrodes . *Appl. Phys. Lett*. **89**, 042505 (2006).

[29] Velev, J. P., Dowben, P. A., Tsymbal, E. Y., Jenkins, S. J. & Caruso, A. N. Interface effects in spin-polarized metal/insulator layered structures. *Surface Science Reports* **63**, 400 (2008).

[30] Caruso, A. N., Schulz, D. L. & Dowben, P. A. Metal hybridization and electronic structure of tris(8-hydroxyquinolato) aluminum (alq(3)). *Chemical Physics Letters* **413**, 321 (2005).

[31] It has been pointed out by the different authors that the real Alq3 thickness involved in the transport is difficult to estimate. According to them, the thickness should be reduced by up to 100 nm, leaving junctions in the few 10nm - 140nm range.

[32] Baldo, M. A. & Forrest, S. R. Interface-limited injection in amorphous organic semiconductors. *Phys. Rev. B* **64**, 085201 (2001).

[33] Nitzan, A. & Ratner, M. Electron transport in molecular wire junctions. *Science* **300**, 1384–1389 (2003).

[34] Mujica, V. Kemp, M. Ratner, M. Electron conduction in molecular wires I A scattering formalism. *J. Chem. Phys.* **101**, 6849-6855 (1994).

[35] Vàzquez, H. et al. Dipole formation at metal/PTCDA interfaces: Role of the Charge Neutrality Level, *Europhys. Lett.* **65**, 802-808 (2004).

[36] Tsymbal, E.Y., Sokolov, A., Sabirianov, I.F., and Doudin, B. Resonant inversion of tunneling magnetoresistance. *Phys. Rev. Lett.* **90**, 186602 (2003).

[37] Garcia, V. et al. Resonant tunneling magnetoresistance in MnAs/III-V/MnAs junctions. *Phys. Rev. B* **72**, 081303 (2005).

[38] Bouzehouane, K. et al. Nanolithography based on real-time electrically controlled indentation with an atomic force microscope for nanocontact elaboration. *Nanoletters* **3**, 1599 (2003).

We acknowledge the financial support from EU-FP6-STRP under Grant No. 033370 OFSPIN and French ANR-PNANO under grant SPINORGA

*Correspondence should be addressed to Pierre Richard. E-mail: pierre.seneor@thalesgroup.com, richard.mattana@thalesgroup.com.

**Figure Legends:**

**FIG. 1: Schematic drawing of the organic magnetic tunnel junction**. The device consists first of a $La_{0.7}Sr_{0.3}MnO_3$/Alq$_3$ bilayer. A nanoindent in the Alq$_3$ layer is performed by a conductive tip AFM allowing us to control the organic tunnel barrier thickness. This nanohole is then filled with cobalt leading to a $La_{0.7}Sr_{0.3}MnO_3$/Alq$_3$/Co nanometric size magnetic tunnel junction.

**FIG. 2: Control of the Alq$_3$ thickness**. **a,** Profile of two nanoindents into Alq$_3$ ( 10nm)/resist( 30nm) bilayer performed at threshold values of $10^{5.5}$ $\Omega$ and $10^{9.5}$ $\Omega$. The $10^{5.5}$ $\Omega$ threshold value corresponds to the AFM tip-LSMO contact resistance, there is no organic material left. On the other hand for a larger threshold value ($10^{9.5}$ $\Omega$) the barrier thickness left is then 4 nm. **b,** Variation of the barrier thickness left as a function of the resistance threshold value (in a log scale) set for the nanoindentation process. The dashed line is a linear fit. The errors bars show the standard deviation obtained on the series of nanoindents.

**FIG. 3: Magnetic field dependence of the resistance**. Magnetoresistance curve of the organic magnetic tunnel junction obtained at 2 K and 10 mV. The lower coercive field corresponds to the LSMO magnetic reversal and the higher coercive field to the Co magnetic switching. In inset I(V) curves recorded at 2 K in the parallel ($I_{PA}$) and antiparallel ($I_{AP}$) magnetic configurations.

**FIG. 4: Bias voltage dependence of the magnetoresistance**. **a,** R(H) magnetoresistance curves of the LSMO/Alq$_3$~2nm/Co nano-MTJ recorded at 2K and at five different bias voltages (-5mV, -10mV, -20mV, -45mV, -100mV). **b,** Bias voltage dependence of the magnetoresistance obtained from I(V) curves recorded in parallel and antiparallel magnetic configurations (line). The data corresponding to the R(H) curves recorded at different bias voltage are also reported (circles) . In inset of b, variation of the magnetoresistance as a function of the temperature.

**FIG. 5: Model used for donor-acceptor mediated transport**. **a,** Schematic drawing of the model of donor-acceptor mediated transport. The donor (resp. acceptor) state is modified by coupling to the left (resp. right) lead are shown in red. Bulk Alq$_3$ transmission between the donor and acceptor is summarized by series of molecular states in black. **b,** Illustration of the interfacial molecular state modification obtained for strong coupling to a ferromagnetic electrode in the limit of $\left| \Gamma \right| >> \left| \Delta \tilde{E} \right|$ with $\left| \Delta \tilde{E} \right| \approx 0$. A simple one band DOS is considered for simplification. The level undergoes a spin dependent broadening while being brought to resonance. Accordingly, its spin polarization at the Fermi level (dot line) is reversed compared to the ferromagnetic electrode one and a new effective spin dependent interface including the first molecular layer has to be defined.

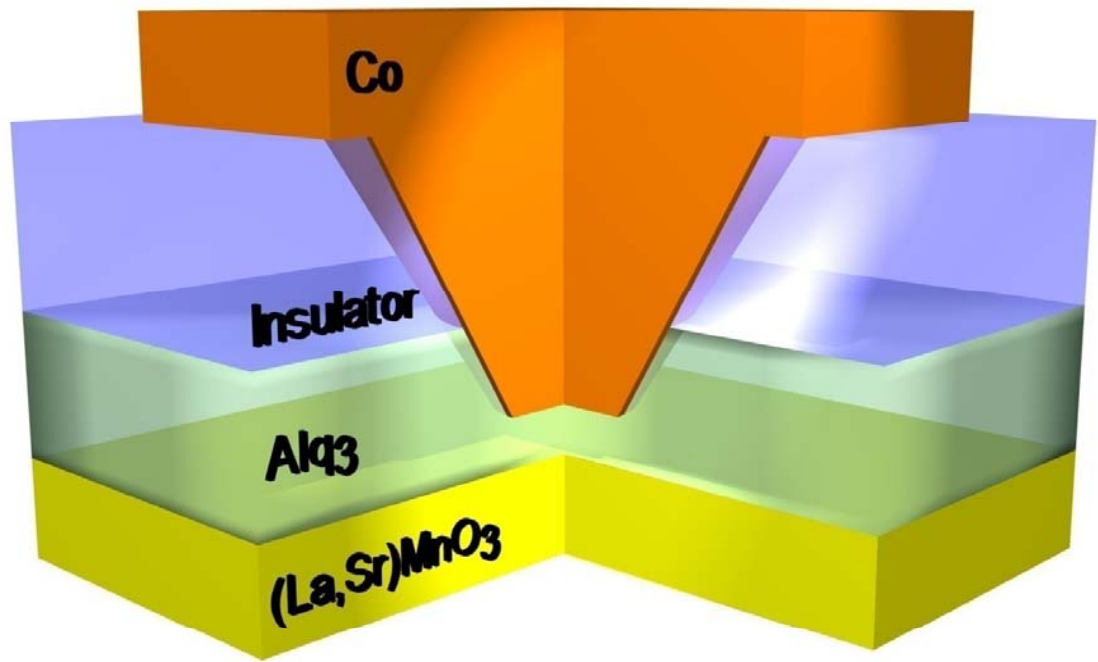

**Figure 1**

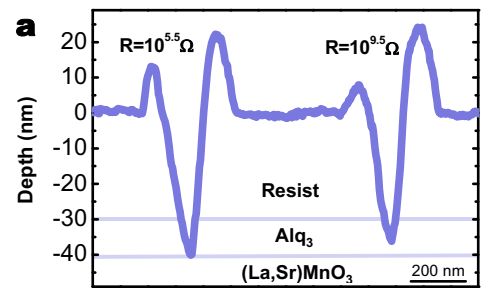

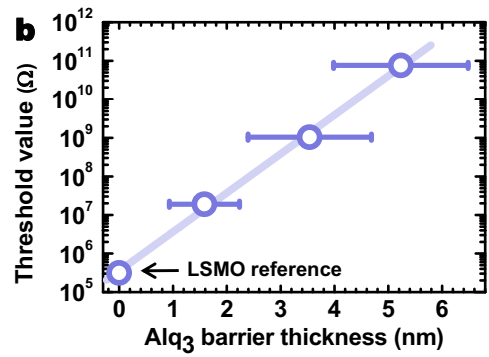

**Figure 2**

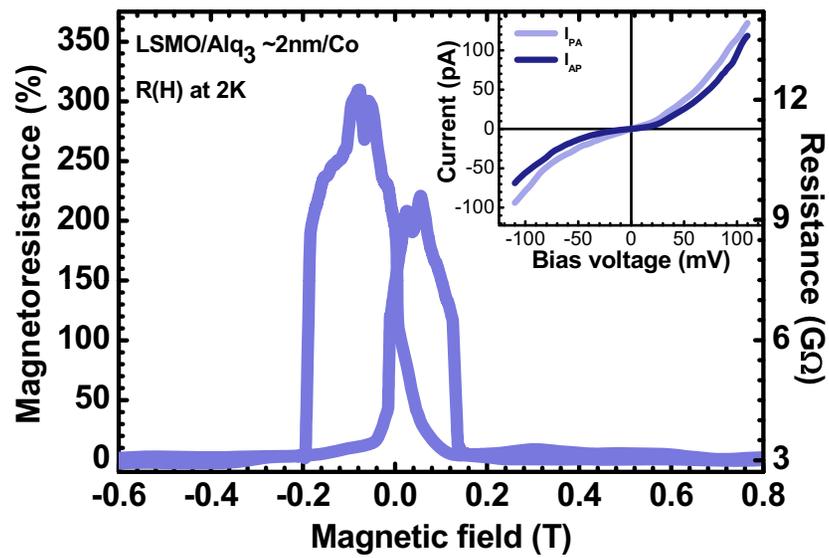

Figure 3

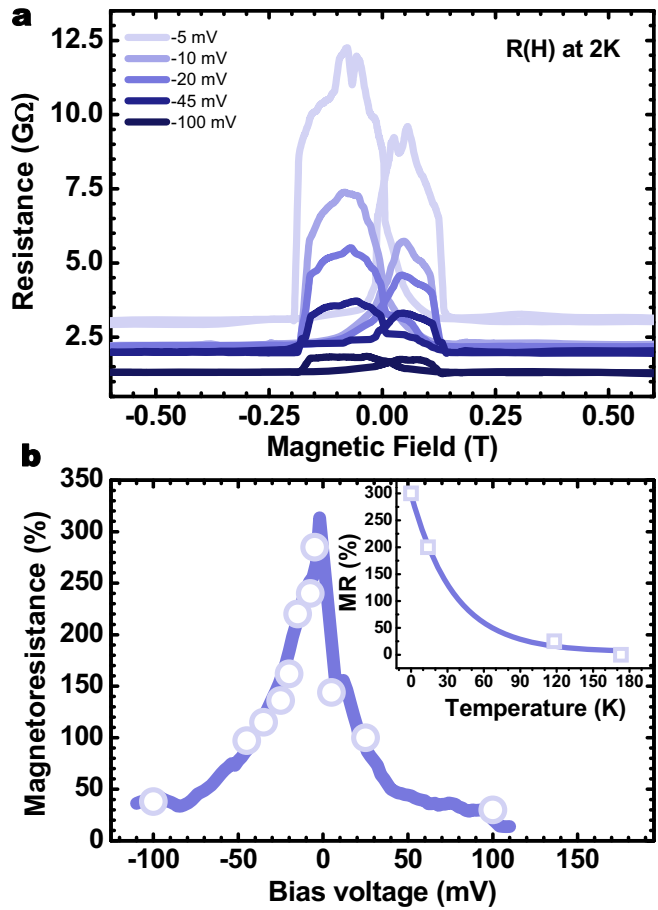

**Figure 4**

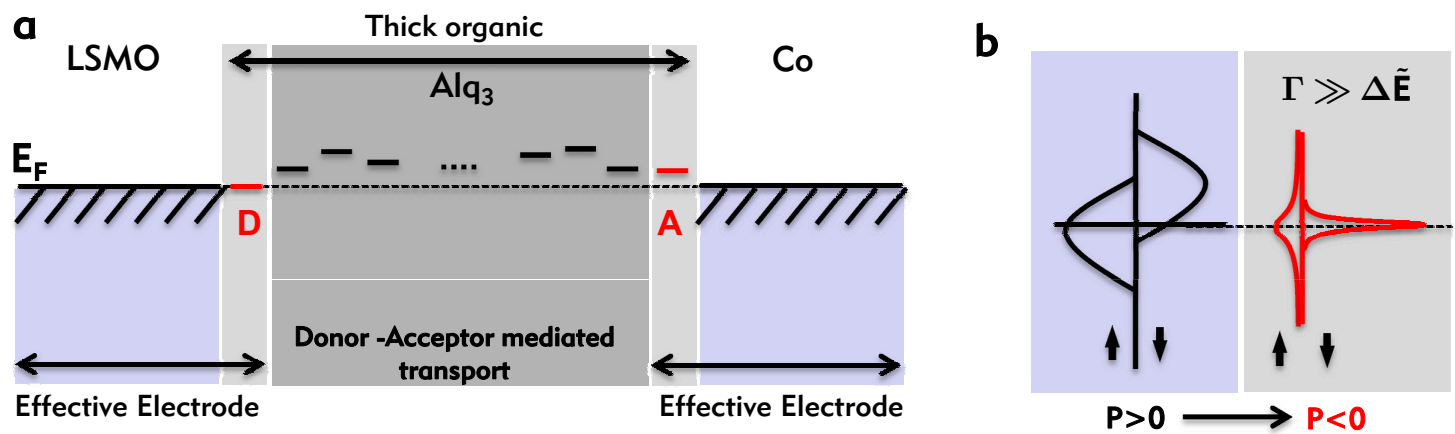

**Figure 5**